\documentclass[aps, prl, twocolumn]{revtex4}
\usepackage{amsmath}
\usepackage{graphicx}
\usepackage{rotating}
\usepackage{bm}
\allowdisplaybreaks
\usepackage{setspace}
\usepackage{titlesec}
\usepackage{units}

\titleformat{\section}[runin]{\normalfont\itshape}{\thesection}{0mm}{}[.---]
\titlespacing{\section}{2mm}{0mm}{0mm}[0mm]

\begin{document}




\title{An exactly size consistent geminal power\\via Jastrow factor networks in a local one particle basis}
\author{Eric Neuscamman\footnote[1]{Electronic mail: eric.neuscamman@gmail.com}}
\affiliation{Department of Chemistry, University of California, Berkeley, California 94720}

\date{\today}

\begin{abstract}
The accurate but expensive product of geminals ansatz may be approximated by a geminal power, but this
approach sacrifices size consistency.
Here we show both analytically and numerically that a size consistent form very similar to the
product of geminals can be recovered using a network of location specific Jastrow factors.
Upon variational energy minimization, the network creates particle number projections that remove the
charge fluctuations responsible for size inconsistency.
This polynomial cost approach captures strong many-electron correlations,
giving a maximum error of just 1.8 kcal/mol during the double-bond dissociation of H$_2$O in an STO-3G
atomic orbital basis.
\end{abstract}

\maketitle


The overwhelming majority of electronic structure methods applied today rely fundamentally on the independent
particle approximation (IPA).
These methods, which include density functional theory \cite{Parr-Yang}, coupled cluster theory
\cite{BARTLETT:2007:cc_review}, configuration interaction \cite{Szabo-Ostland}, and many body perturbation
theory \cite{Szabo-Ostland}, all assume that the wave function is well approximated by a single Slater
determinant (SD) in which the only correlations between electrons are those due to Fermi statistics.
This assumption fails dramatically in a number of important cases displaying \textit{strong correlation}
between electrons, including multiple-bond breaking, excited states, transition metal compounds, and lattice
Hamiltonians used in the study of high temperature superconductivity.
While this failure can in some cases be rectified by active space methods that employ linear combinations of 
determinants, these methods' costs increase exponentially with system size.
Indeed, when developing methods to treat strong correlation, one prefers to retain the formal
properties of the SD: polynomially scaling cost, energies that are variational (i.e.\ upper bounds),
and size consistency, in which two non-interacting systems give the same total energy when modeled separately
or together.

One approach to this ideal is to generalize the SD, which is a product of one-particle
functions (orbitals), to a product of two-particle functions (geminals), known as the antisymmetric product of geminals (APG).
\begin{align}
\label{eqn:apg}
|\Psi_{\mathrm{APG}}\rangle = \prod_{i=1}^{N/2} \hat{G}_i |0\rangle \quad \quad \hat{G}_i = \sum_{r s} g^i_{r s} a^\dag_{r\uparrow} a^\dag_{s\downarrow}
\end{align}
Here each operator $\hat{G}_i$ creates a pair of opposite-spin electrons in a two-particle geminal defined by the
weights $g^i_{r s}$ and operators $a^\dag_{r\uparrow}$ and $a^\dag_{s\downarrow}$ that create $\uparrow$ and $\downarrow$
electrons in the sites (or orbitals) $r$ and $s$.
(The conclusions in this Letter generalize to same-spin pairs and pfaffians \cite{Bajdich:2006:pfaffian,Bajdich:2008:pfaffian},
but to avoid unnecessary complication we restrict ourselves to opposite-spin pairs.)
If no restrictions are placed on the form of the geminals, the resulting wave function has been shown to be highly
accurate. \cite{Nicely:1971:apg,Silver:1969:apg}
However, the author is not aware of any polynomial cost, variational methods for working with the general APG, and
indeed it is more often approximated by requiring that the geminals be built from separate, mutually orthogonal
sets of one-particle functions (APSG) \cite{Kutzelnigg:1964:apsg,Kutzelnigg:1965:apsg,Surjan:2012:apsg},
resulting in methods such as perfect pairing (PP) \cite{POPLE:1953:agp}
and the resonating valence bond (RVB) \cite{Goddard:1972:rvb}.
While APSG methods can achieve size consistency, variational energies, and polynomial cost, they lack
correlation between electron pairs \cite{Kutzelnigg:1999:cumulants} and are thus unsuitable for treating strong
correlations between more than two electrons \cite{Parkhill:2009:perfect_quads}.
While corrections can be applied via configuration interaction
\cite{Goddard:1975:gvbci,Surjan:2012:apsg}, coupled cluster \cite{Parkhill:2009:perfect_quads,Parkhill:2010:perfect_hexs,Small:2009:ccvb},
perturbation theory \cite{Surjan:2012:apsg}, and Hopf algebra \cite{Cassam:2003:hopf_alg,Cassam:2006:hopf_alg},
none of these approaches
simultaneously retain polynomial cost, variational energies, and size consistency.

Building on the work of Casula and Sorella
(see Refs.\ \cite{Sorella:2003:agp_sr,Sorella:2004:agp_sr,Sorella:2009:jagp_molec} and especially \cite{Sorella:2007:jagp_vdw}),
we present an ansatz that captures strong inter-pair correlations while retaining
polynomial cost, variational energies, and size consistency.
To the best of our knowledge, this is the first example of a method that achieves all of these properties
for a general system and an ab initio Hamiltonian.

\section{Ansatz}
\label{sec:ansatz}

We begin our construction with the well-known geminal power (AGP) ansatz,
\begin{align}
\label{eqn:agp}
|\Psi_{\mathrm{AGP}}\rangle = \hat{F}^{N/2} |0\rangle \quad \quad \hat{F} = \sum_{r s} f_{r s} a^\dag_{r\uparrow} a^\dag_{s\downarrow} \simeq \sum_i \hat{G}_i,
\end{align}
in which the (bosonic) electron pairs all reside in the same low-energy geminal $\hat{F}$,
which should be similar to the sum of the APG geminals $\hat{G}_i$.
For those more familiar with the superconducting Bardeen-Cooper-Schrieffer (BCS) ansatz \cite{Bardeen:1957:bcs},
it may be helpful to consider that the AGP is the RVB equivalent of a particle-number-projected BCS, with the
real space pairing matrix $f_{r s}$ related by a Fourier transform to the BCS $k$-space weights
(see Ref.\ \cite{Ogata:2008:tj_model}, Eqs.\ 4.9-4.10).
While the AGP admits a number of polynomial cost, variational methods
\cite{Kurtz:1982:gagp,Ortiz:1981:gagp,Weiner:1980:gagp,Scuseria:2002:hfb,Sorella:2003:agp_sr,Sorella:2004:agp_sr,Scuseria:2011:projected_hfb}
(one of which \cite{Scuseria:2011:projected_hfb} achieves a mean-field $n^3$ cost), it suffers
from a severe size consistency problem resulting from terms in which a single operator $\hat{G}_i$ is repeated, placing
four or more electrons in the same local geminal.

Our approach is to eliminate these charge transfer or ``ionic'' terms by
enforcing local particle number distributions with a network of Jastrow factors.
We thus produce a size consistent Jastrow-AGP (JAGP) ansatz,
\begin{align}
\label{eqn:jagp}
|\Psi_{\mathrm{JAGP}}\rangle = \exp\left( \sum_{p q} \sum_{\sigma \tau \in \uparrow,\downarrow} \hat{J}_{p_\sigma q_\tau} \right) \hat{F}^{N/2} |0\rangle,
\end{align}
in which the Jastrows operate on the bare AGP in the same way they operate on a bare SD in the traditional Jastrow-Slater ansatz \cite{FouMitNeeRaj-RMP-01}.
The Jastrow factors $\hat{J}_{p_\sigma q_\tau}$ inspect the occupation ($00$, $01$, $10$, or $11$) of each orbital pair
$p_\sigma q_\tau$ and apply a corresponding scalar factor to the wave function to favor or penalize the configuration of that particular pair.
They are defined by
\begin{align}
\label{eqn:jf}
\hat{J}_{p_\sigma q_\tau} = \sum_{n,m\in 0,1} C^{p_\sigma q_\tau}_{n m} \hat{P}^{p_\sigma}_{n} \hat{P}^{q_\tau}_{m},
\end{align}
where $\bm{C}$ is a tensor of penalty factors and the operator $\hat{P}^{p_\sigma}_{n}$ gives one if orbital $p_\sigma$ has occupation $n$
and zero otherwise.
Note that together, these Jastrow factors are equivalent to the correlator product state tensor network
\cite{CHANGLANI:2009:cps,MEZZACAPO:2009:entangled_plaquettes},
and so we refer to them as a location-specific Jastrow factor network.
In this Letter, we demonstrate that this network restores size consistency to the
AGP by imposing local particle number constraints,
delivering an ansatz that is similar in character to the APG, size consistent, variational, polynomial cost, and effective at
treating strong many-electron correlations.

\begin{figure}[t]
\centering
\includegraphics[width=6.0cm,angle=0]{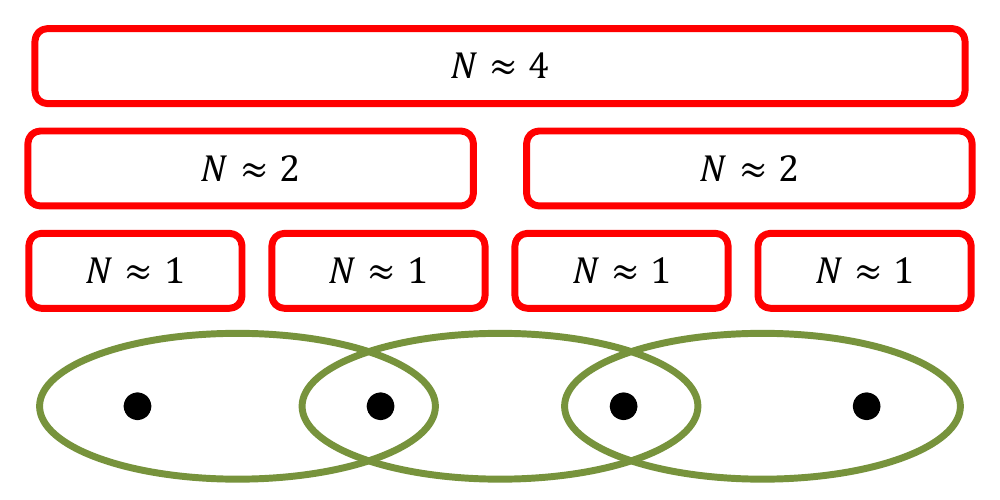}
\caption{(color online) A cartoon schematic of our JAGP ansatz, in which the geminal power is constructed
         from non-orthogonal local geminals (ovals) describing bonds between neighboring atoms (circles).
         Particle number projectors (rectangles) built from Jastrow factors constrain electron counts on
         atoms and groups of atoms to remove the ``ionic'' AGP terms responsible for size consistency errors.}
\label{fig:schematic}
\end{figure}

\section{Charge fluctuations}
\label{sec:charge_transfer}

To be size consistent, a wave function must factor into a product of subsystem
wave functions $|\Psi_{AB}\rangle=|\Psi_{A}\rangle|\Psi_{B}\rangle$ when applied to two non-interacting
subsystems $A$ and $B$.
As noted previously by Sorella, Casula, and Rocca \cite{Sorella:2007:jagp_vdw}, unphysical charge
fluctuations prevent this factorization.
For a simple example, imagine two H$_2$ molecules described by geminals $\hat{G}_A$ and $\hat{G}_B$.
The AGP built from these geminals, $(\hat{G}_A + \hat{G}_B)^2|0\rangle$, contains both the
neutral term $\hat{G}_A \hat{G}_B |0\rangle$ and the unphysical ionic terms $\hat{G}_A^2|0\rangle$
and $\hat{G}_B^2|0\rangle$ in which all four electrons reside on one molecule.
Without the ionic terms, this AGP would factor correctly and be size consistent.

We may generalize this analysis by expanding the AGP in the basis of occupation number vectors
$|\bm{n}\rangle$, each of which specifies a unique occupation pattern of the orbitals.
\begin{align}
\label{eqn:on_agp}
|\Psi_{\mathrm{AGP}}\rangle = \sum^{N_\uparrow N_\downarrow}_{\bm{n}} \det \bm{\Phi}_{\bm{n}} |\bm{n}\rangle
\end{align}
Here the coefficients simplify to determinants of the occupied pairing matrices $\bm{\Phi}_{\bm{n}}$
\cite{Bouchaud:1988:agp_det}, which are obtained by deleting from $\bm{f}$ rows and columns
corresponding to unoccupied orbitals.
Note that the sum is restricted to states with the correct total $\uparrow$ and $\downarrow$ electron counts
$N_\uparrow=N_\downarrow=N/2$.

An intuitive guess for $|\Psi_{AB}\rangle$ is to take the AGP geminal as the sum of the subsystem geminals
and the Jastrow factor as the product of the subsystem Jastrows, in which case the pairing matrix $\bm{f}$
will be block diagonal with blocks equal to the subsystem matrices $\bm{f}_A$ and $\bm{f}_B$, and the
Jastrows will be defined by $\bm{C}=\bm{C}_A+\bm{C}_B$.
Such a choice results in
\begin{align}
\label{eqn:psi_ab}
|\Psi_{A B}\rangle = e^{\hat{J}_A}e^{\hat{J}_B}\sum^{N_\uparrow N_\downarrow}_{\bm{n} = \bm{n}_A,\bm{n}_B} \det \bm{\Phi}_{\bm{n}_A} \det \bm{\Phi}_{\bm{n}_B} |\bm{n}\rangle,
\end{align}
where the determinant factors due to the block-diagonality of $\bm{f}$ and the Jastrow
factors due to the additive separability of $\bm{C}$.
However, $|\Psi_{A B}\rangle$ \textit{does not} factor into $|\Psi_{A}\rangle|\Psi_{B}\rangle$,
because the summation over orbital occupations $\bm{n}$ contains ionic terms in which electrons are
transferred between subsystems.

Using real space three-body Jastrow factors, Sorella et al showed \cite{Sorella:2007:jagp_vdw} that these
spurious charge fluctuations can be partially suppressed, mitigating the size consistency error.
However, removing the error completely through this approach would require unlimited flexibility in the Jastrow.
In practice, their wave function retained a size consistency error on the order of 1eV in the carbon dimer
\cite{Sorella:2007:jagp_vdw}, although the effect on binding energies was much smaller due to error cancellation.
We expand on this idea, showing how Jastrow factors can eliminate the size consistency error entirely.

\section{Partial number projection}
\label{sec:np}

Consider the operator
\begin{align}
\label{eqn:part_proj_op}
\hat{Q}(\alpha,M,X) = \exp\left(-\alpha\left(M-\sum_{p\in X}\hat{P}^{p}_1\right)^2\right),
\end{align}
which we call a partial number projection operator favoring $M$ electrons in the set of orbitals $X$.
In the limit $\alpha\rightarrow\infty$, this becomes a strict projection, deleting terms
in which $X's$ electron count differs from $M$.
We may thus fix the subsystem electron counts and delete ionic terms using the operators
$\hat{Q}_A = \hat{Q}(\alpha,N_{A\uparrow},A_\uparrow) \hat{Q}(\alpha,N_{A\downarrow},A_\downarrow)$
and
$\hat{Q}_B = \hat{Q}(\alpha,N_{B\uparrow},B_\uparrow) \hat{Q}(\alpha,N_{B\downarrow},B_\downarrow)$,
which when applied to Eq.\ (\ref{eqn:psi_ab}) produce the desired factorization.
\begin{align}
\label{eqn:proj_ab}
\lim_{\alpha \to \infty}\hat{Q}_A\hat{Q}_B|\Psi_{A B}\rangle &= |\Psi_A\rangle |\Psi_B\rangle \\
\notag  |\Psi_A\rangle&=e^{\hat{J}_A} \sum^{N_{A\uparrow} N_{A\downarrow}}_{\bm{n}_A} \det \bm{\Phi}_{\bm{n}_A} |\bm{n}_A\rangle \\
\notag  |\Psi_B\rangle&=e^{\hat{J}_B} \sum^{N_{B\uparrow} N_{B\downarrow}}_{\bm{n}_B} \det \bm{\Phi}_{\bm{n}_B} |\bm{n}_B\rangle
\end{align}
Thus if we can apply appropriate projections, JAGP will factor and be size consistent.

The advantage of our ansatz is that the partial projection operators $\hat{Q}_A$ and $\hat{Q}_B$
can be built into the Jastrow network.
To see how, expand the square in Eq.\ (\ref{eqn:part_proj_op}) and drop the constant term
$\exp(-\alpha M^2)$, which only affects normalization, to obtain
\begin{align}
&\hat{Q}(\alpha,M,X) \notag \\
&\quad\rightarrow \exp\left(2M\alpha\sum_{p\in X}\hat{P}^{p}_1 - \alpha \sum_{p,q\in X}\hat{P}^{p}_1\hat{P}^{q}_1\right)
\label{eqn:expand_square} \\
&\quad= \exp\left(\sum_{p,q\in X}\beta\hat{P}^{p}_1\hat{P}^{q}_0 + \left(\beta-\alpha\right)\hat{P}^{p}_1\hat{P}^{q}_1\right),
\label{eqn:rearranged_square}
\end{align}
where $\beta=2M\alpha/k$ and $k$ is the number of orbitals in $X$.
Inspecting Eq.\ (\ref{eqn:rearranged_square}) reveals that the Jastrow network
defined in Eqs.\ (\ref{eqn:jagp}) and (\ref{eqn:jf}) can contain any combination
of partial projection operators.
The JAGP is therefore capable of deleting ionic terms by restricting subsystem 
electron counts, making it factorizable and size consistent.
Furthermore, if we take our AGP geminal as a sum of the localized but non-orthogonal
APG geminals, partial number projections can help ensure that each local geminal has
the correct number of electrons.
Our JAGP thereby emulates the structure of the APG.

\begin{figure}[t]
\centering
\includegraphics[width=7.0cm,angle=0]{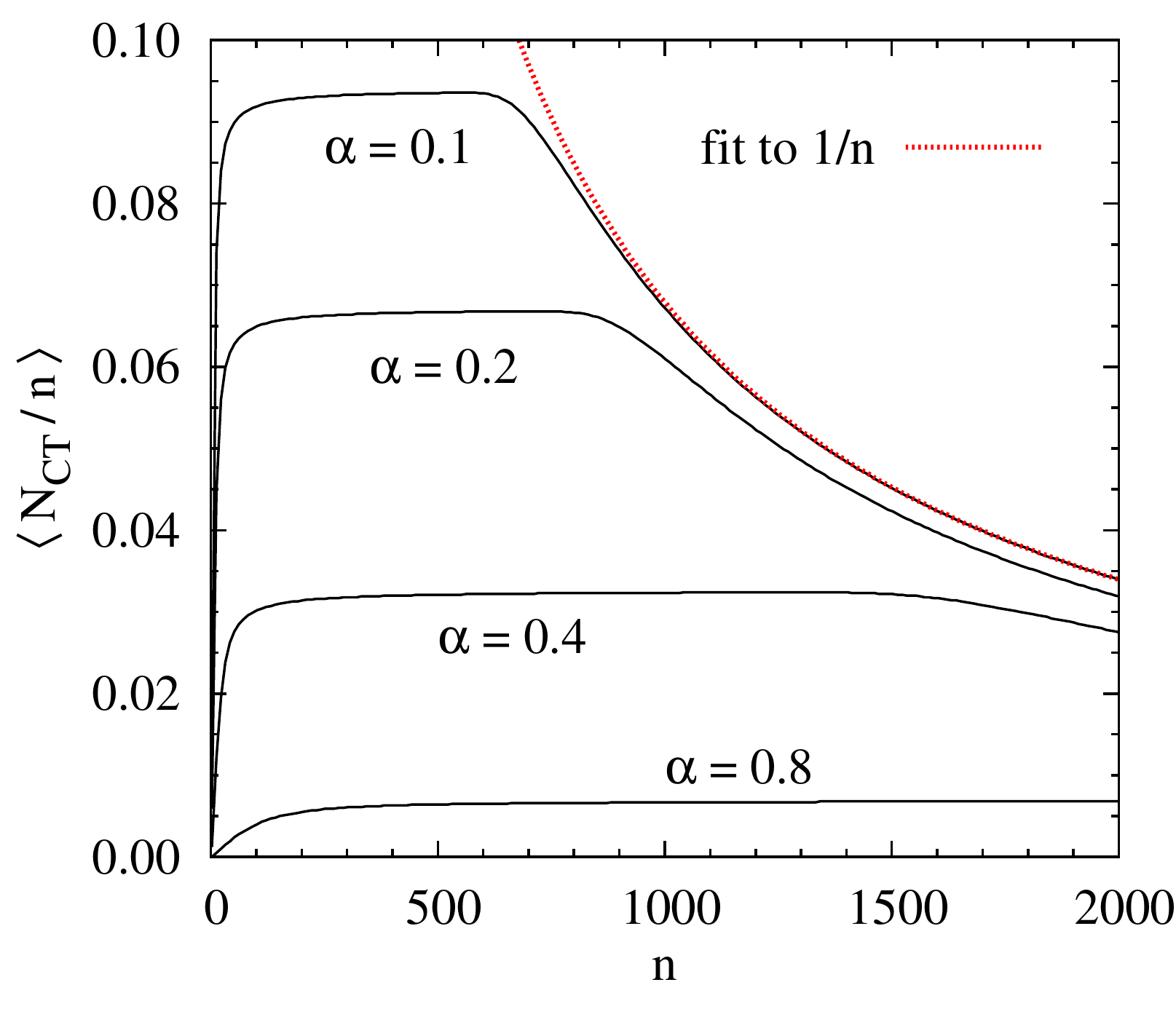}
\caption{(color online) The average number of unphysical charge transfers per molecule in a
         system of $n$ well separated H$_2$ molecules.
         The wave function is a PP-parameterized AGP with various partial number projections.
         The dotted line is a fit showing the asymptotic $1/n$ decay for $\alpha=0.1$.}
\label{fig:h2_ct}
\end{figure}

\begin{figure}[t]
\centering
\includegraphics[width=7.0cm,angle=0]{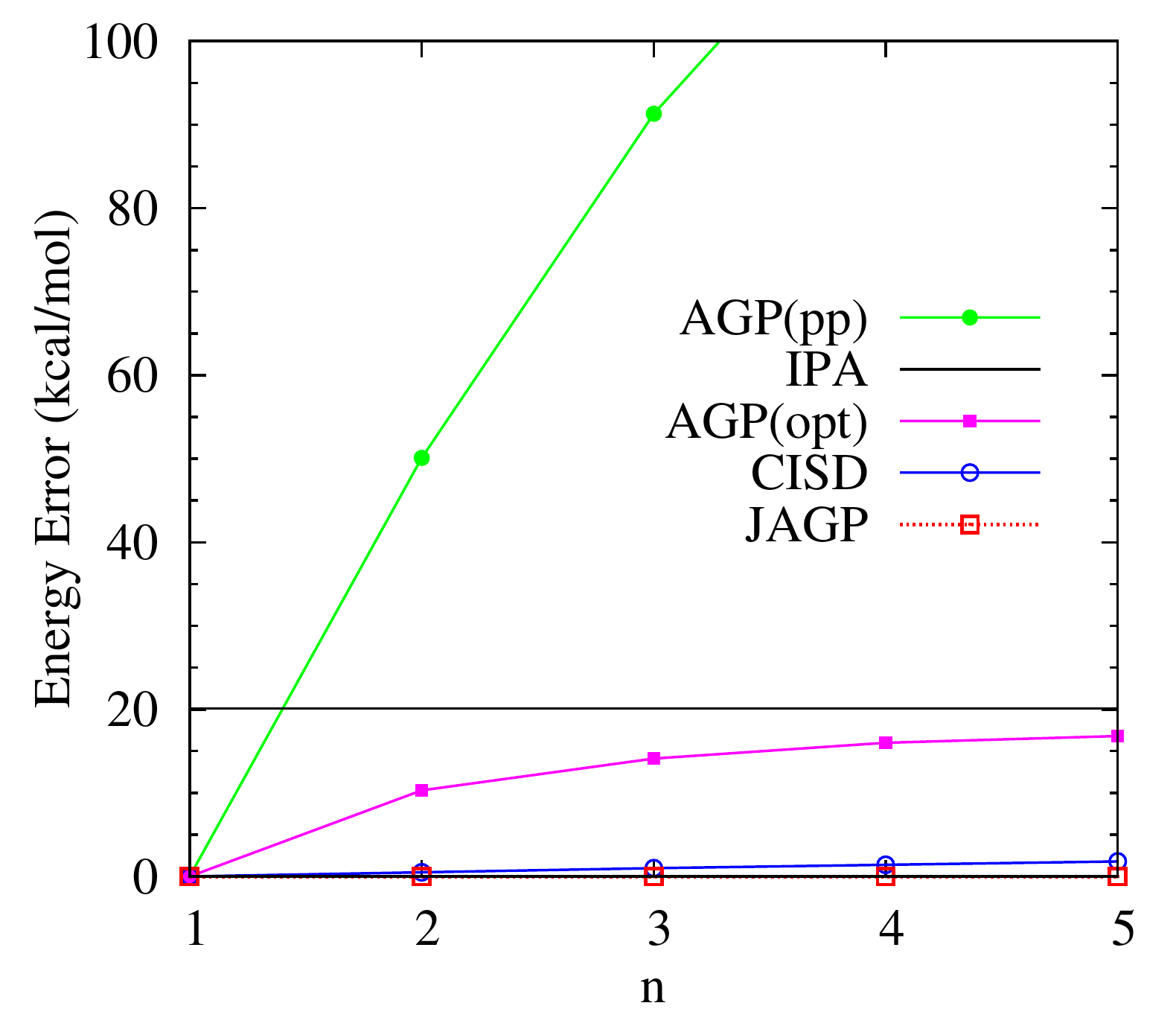}
\caption{(color online) Energy errors per molecule for $n$ well-separated H$_2$ molecules.
         For AGP, both the PP and optimized versions of the wave function are shown.}
\label{fig:h2_error}
\end{figure}

\section{Variational minimization}
\label{sec:opt}

We use variational Monte Carlo (VMC) \cite{FouMitNeeRaj-RMP-01,NigUmr-BOOK-99} to evaluate and minimize
the JAGP energy by varying independently all elements of the pairing matrix $\bm{f}$ and Jastrow factor
penalty tensor $\bm{C}$.
The Hamiltonian is the typical Born-Oppenheimer approximation to the electronic Hamiltonian with
relativistic terms neglected.
Note especially that we work in Fock space rather than real space.
We use an improved version of the Linear Method optimizer along the lines proposed in
Ref.\ \cite{Neuscamman:2012:fast_sr},
the details of which will be presented elsewhere \cite{Neuscamman:xxxx:fast_lm}.
For the present discussion, it suffices to convey that this method is variational with a cost of
$O(n_s n_o^2 n_u^2)$, where $n_s$, $n_o$, and $n_u$ are the sample size and the numbers
of occupied and unoccupied orbitals.

\section{Hydrogen gas}
\label{sec:h2}

A collection of $n$ well separated hydrogen molecules reveals the severity of AGP's charge fluctuations.
Working in a symmetrically orthogonalized STO-3G basis \cite{Pople:1969:sto-3g}, in which a single 1s orbital is centered
on each H, we may define the AGP geminal as a sum of PP geminals,
\begin{align}
\label{eqn:pp_sum}
|\Psi_{n\mathrm{H_2}}\rangle = \hat{Q}\left( \sum_i^n x g^\dag_{i\uparrow}g^\dag_{i\downarrow} + y u^\dag_{i\uparrow}u^\dag_{i\downarrow} \right)^n |0\rangle.
\end{align}
Here $x^2+y^2=1$, $\hat{Q}$ is a partial number projection operator suppressing charge fluctuations, and
$g^\dag_{i\uparrow/\downarrow}$ and $u^\dag_{i\uparrow/\downarrow}$ create electrons in the bonding
and antibonding orbitals, respectively, of the $i$th H$_2$ molecule.
If we parameterize $\hat{Q}$ to apply a penalty of $e^{-2\alpha}$ for each incorrect H$_2$ electron
count, then the average number of charge transfers (i.e.\ the number of $[$H$_2]^{2+}$
ions) will be
\begin{align}
\label{eqn:avg_n_ct}
\langle N_{CT} \rangle =
\frac{\sum_{l=0}^{n/2} ~ l e^{-8\alpha l}\left(\frac{x^{l}y^{l}}{l!}\right)^2 \frac{n!}{(n-2l)!}}{\sum_{l=0}^{n/2} ~ e^{-8\alpha l}\left(\frac{x^{l}y^{l}}{l!}\right)^2 \frac{n!}{(n-2l)!}} ~,
\end{align}
where the contributions are grouped by the number of charge transfers $l$.
Figure \ref{fig:h2_ct} shows that $\langle N_{CT}/n \rangle$ decays as
$1/n$ in the thermodynamic limit $n\rightarrow\infty$, recovering the
size extensivity of the BCS ansatz.
However, the steep growth of $\langle N_{CT}/n \rangle$ for small $n$ is unacceptable for quantum
chemistry, where system sizes range from tens to hundreds of bonding electron pairs.
Encouragingly, $\langle N_{CT}/n \rangle$ is very sensitive to increasing $\alpha$,
showing that charge fluctuations are easily suppressed in our ansatz.

\begin{figure}[t]
\centering
\includegraphics[width=7.0cm,angle=0]{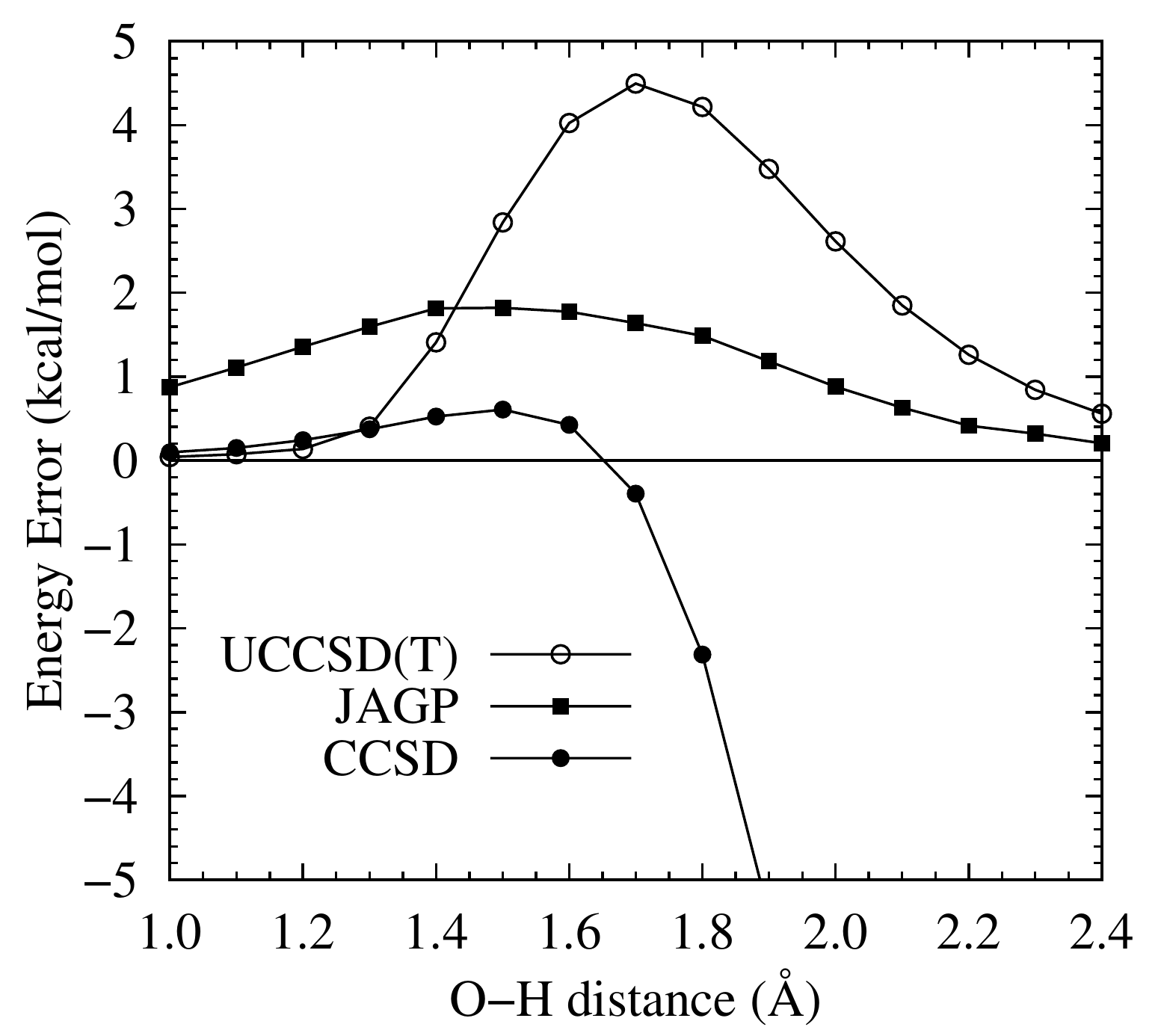}
\caption{Energy errors relative to FCI for the symmetric dissociation of minimal basis H$_2$O
         with bond angle 109.57$^\circ$.
         JAGP's statistical uncertainties are smaller than the symbols.}
\label{fig:h2o_error}
\end{figure}

Without partial number projection, the charge fluctuations in an AGP built from PP geminals
render it less accurate than the IPA for small $n$, as shown in Figure \ref{fig:h2_error}.
Here we use the somewhat more realistic symmetrically orthogonalized 6-31G basis \cite{POPLE:1972:6-31g_basis}.
The errors for the variationally optimized AGP are less than those of the IPA, but they remain
large compared to those of singles and doubles configuration interaction (CISD),
whose well-known size consistency problem is much less severe.
Most importantly, variational optimization (with initial guess $\bm{f}=\mathrm{random}$, $\bm{C}=0$)
of our JAGP completely removes size consistency errors and produces the exact PP result.
This shows our optimization can discover the need for particle number projection and
impose it automatically.

It is worth noting that a significant component of the JAGP's correlation energy is size extensive
(i.e.\ it scales linearly with system size for large systems), because the JAGP can always be made to
contain PP, and PP energies are size extensive.
Less clear is whether the entire JAGP energy is extensive, which clearly merits further investigation.

\begin{table}[t]
\centering
\caption{
         JAGP energies for collections of $n$ well separated H$_2$O molecules
         with bond lengths 1.4\AA\ and angles 109.57$^\circ$.
         Statistical uncertainty in final digit given in parentheses.
        }
\label{tab:multi_h2o}
\begin{tabular}{  r   r@{.}l  }
\hline\hline
\multicolumn{1}{ c }{ \hspace{0mm} $n$            \hspace{0mm} } &
\multicolumn{2}{ c }{ \hspace{0mm} $E/n$ (a.u.)   \hspace{0mm} } \\
\hline
 \hspace{3mm} 1 \hspace{3mm} & \hspace{2mm} -74&90371(1) \hspace{2mm} \\
 \hspace{3mm} 2 \hspace{3mm} & \hspace{2mm} -74&90374(3) \hspace{2mm} \\
 \hspace{3mm} 4 \hspace{3mm} & \hspace{2mm} -74&90369(3) \hspace{2mm} \\
 \hspace{3mm} 8 \hspace{3mm} & \hspace{2mm} -74&90376(5) \hspace{2mm} \\
\hline\hline
\end{tabular}
\end{table}

\section{Double bond dissociation}
\label{sec:double_bond}

To demonstrate JAGP's ability to capture strong inter-pair correlations while maintaining
size consistency, we have applied it to the symmetric double-bond dissociation of H$_2$O in a symmetrically
orthogonalized STO-3G basis.
We first optimized the wave function for a single molecule, starting from a very poor initial guess
($\bm{f}=\mathrm{random}$, $\bm{C}=0$).
Figure \ref{fig:h2o_error} shows that the maximum error relative to full configuration interaction
(FCI) is 1.8 kcal/mol, a factor of 2.5 smaller than the 4.5 kcal/mol error produced by
unrestricted coupled cluster with singles, doubles, and perturbative triples (UCCSD(T)).
In terms of correlation energies (defined with respect to an unrestricted SD), JAGP retains above
90\% of the correlation across the whole curve, while UCCSD(T) is less balanced with correlation recovery
ranging from over 99\% near equilibrium down to 75\% upon dissociation.

After optimizing our ansatz for one water molecule, we tested size consistency by constructing wave
functions for two, four, and eight well separated water molecules.
The overall geminals were built as sums of monomer geminals, and the Jastrow
tensor $\bm{C}$ as the sum of the monomers' plus the terms necessary to impose partial number projection
with $\alpha=2$ on the $\uparrow$ and $\downarrow$ electron occupations of each molecule.
Table \ref{tab:multi_h2o} reveals that the energy per molecule is independent of the number of molecules,
showing that JAGP is size consistent even when it is not exact (as was the case for H$_2$).
Finally, note that the eight water case corresponds to a 40-orbital active space.

\section{Conclusions}
\label{sec:conclusions}

We have shown that a geminal power augmented with a network of location-specific Jastrow factors recovers
size consistency in a localized one particle basis, producing an ansatz similar in character to the
powerful but expensive product of non-orthogonal geminals.
The resulting method is variational, size consistent, polynomial cost, and able to capture strong
many-electron correlations.
It completely removes unphysical charge fluctuations from a dilute H$_2$ gas and accurately
captures the strong correlations of water's double-bond dissociation.
We believe it is the first geminal method satisfying all of these properties and that it is
a promising candidate for applications to other strongly correlated systems.


We thank Martin Head-Gordon for helpful conversations and computational resources.
We thank the Miller Institute for Basic Research in Science for funding.


\bibliography{size_consistent_jrgp.bbl}

\end{document}